\documentclass[11pt,letterpaper]{article}
\usepackage{emnlp2017}
\usepackage{times}
\usepackage{latexsym}
\usepackage{url}
\usepackage{enumitem}
\usepackage{amsmath}
\usepackage{graphicx}
\usepackage{amssymb}
\usepackage{amsfonts}
\usepackage{bm}
\usepackage{floatrow}
\usepackage{latexsym}
\usepackage{makecell}

\newfloatcommand{capbtabbox}{table}[][\FBwidth]

\usepackage[show]{notes}

\usepackage{makecell}	
\usepackage{multirow}	

\newcommand{\ignore}[1]{}

\newcommand{\com}[1]{}

\newcommand{\lisa}[1]{\inote[LISA]{\textcolor{red}{\bf #1 }}}

\def\citeapos#1{\citeauthor{#1} (\citeyear{#1})}

\emnlpfinalcopy

\def\emnlppaperid{***}

\title{ConStance: Modeling Annotation Contexts to Improve Stance Classification}

\author{Kenneth Joseph\textsuperscript{1}~~ Lisa Friedland\textsuperscript{1} ~~ Will Hobbs\textsuperscript{1} ~~ David 
Lazer\textsuperscript{1} ~~ Oren Tsur\textsuperscript{1,2} \\
  {\tt \{k.joseph, l.friedland,  w.hobbs\}@northeastern.edu} \\
  {\tt d.lazer@neu.edu}, {\tt orentsur@bgu.ac.il } \\
  \textsuperscript{1}Network Science Institute \qquad  \textsuperscript{2}Software and Information Systems Engineering \\ 
Northeastern University ~~~~~~~~~~~~~~\qquad Ben Gurion University of the Negev~~~~~ }

\date{}

\begin{document}

\maketitle

\begin{abstract}
Manual annotations are a prerequisite for many applications of machine learning.
However, weaknesses in the annotation process itself are easy to overlook. In particular, scholars often choose what information to give to annotators without examining these decisions empirically.
For subjective tasks such as sentiment analysis, sarcasm, and stance detection, such choices can impact results. 
Here, for the task of political stance detection on Twitter, we show that providing
too little context can result in noisy and uncertain annotations, 
whereas providing too strong a context may cause it to outweigh other signals.
To characterize and reduce these biases, we develop ConStance, a general model for reasoning about annotations across information 
conditions. 
Given conflicting labels produced by multiple annotators seeing the same instances with different contexts, ConStance simultaneously 
estimates gold standard labels and also learns a classifier for new instances. We show that the classifier learned by ConStance outperforms 
a variety of baselines at predicting political stance, while the model's interpretable parameters shed light on the effects of each context.

\end{abstract}

\section{Introduction}
\label{sec:intro}

When annotators are asked for objective judgments about a text (e.g., POS tags), the broader context in which the text is situated is often 
irrelevant. However, many NLP tasks focus on inference of factors beyond words and syntax.  For example, the present work addresses the 
task of detecting political stance on Twitter. We ask annotators to determine whether a given Twitter user supports Donald Trump or Hillary 
Clinton. 
However, inferring something about a \emph{user} from a single tweet that she writes may prove difficult. Prior work on stance has relied on annotations collected this way \citep{mohammad_stance_2016}, but individual tweets do not always contain clear indicators.

%
%
%


One solution to this issue is to supply the annotator with more information about the user. For example, for the similar task of classifying a Twitter user's political affiliation, \citeapos{cohen_classifying_2013-1} display the user's last 10 tweets.
\citet{nguyen_how_2013}, studying gender and age, ask annotators to label users by leveraging all information available in their profile. Thus, researchers have provided a range of contexts (or more broadly, information conditions) to annotators in an attempt to balance annotators' exposure to the data 
needed for accuracy
with reasonable costs in terms of time, money and cognitive load.

However, while scholars routinely make such decisions about what information to show annotators, they rarely examine how such decisions actually impact annotations.  The first contribution of this paper (Section~\ref{sec:indiv_contexts}) is to show that, at least for political stance detection on Twitter, 
displaying different kinds of context to annotators yields significantly different annotations \emph{for the same user}.
\ignore{\lisa{**}} As a result of these discrepancies, the accuracy of models trained on these annotations varies widely.


While it is possible one could select a ``best'' context for a given task, our results suggest that doing so \emph{a priori} is difficult and that, moreover, different contexts provide complementary information. What we would prefer, instead, is a model that \emph{learns} how contexts affect annotators and \emph{combines} annotations from multiple contexts to create gold standard labels. 

Fortunately, prior work suggests mechanisms for such a model.  Typically in annotation tasks, each item is judged by several annotators, and the resulting labels are aggregated, usually by majority vote, to create a gold standard. As an alternative to majority vote, \citet{raykar_learning_2010} develop an elegant probabilistic approach for learning to aggregate labels produced by annotators of varying quality. Their model jointly estimates gold standard labels (in the form of probability scores), infers annotator error rates, and learns a classifier for use on out-of-sample data.




Our second contribution (Section~\ref{sec:model}) is an extension of Raykar et al.'s model to handle labels not only created by annotators
of varying quality, but also produced under \emph{information conditions} of varying quality. We call this model 
ConStance\footnote{
Replication materials for this work, including code for ConStance, are available at 
\url{https://github.com/kennyjoseph/constance}. The paper's Supplementary Material can also be accessed there.
}.  Like \citet{raykar_learning_2010}, who find that even low-quality annotators are 
useful, we find that low-quality contexts can be useful.  
Specifically, we find that the classifier produced from our model performs better than any classifier trained by majority vote from the same labels. Furthermore, the model provides an unsupervised method for comparing the information conditions by examining their respective error patterns.



Intuitively, ConStance performs a role analogous to boosting for annotations: for an arbitrary task, it permits collection of labels that capture different aspects of the instances at hand, then combines them automatically to determine which are more reliable and to produce a classifier that takes all this into account.

\section{Annotating Political Stance}
\label{sec:stance}
\subsection{Political Stance Detection}

Stance detection is defined as the task of determining whether an individual is in favor of, against, or neutral 
towards a target concept based on the content they have generated \citep{mohammad_stance_2016}. It is related to but 
distinct from sentiment analysis: a given document can have negative sentiment but a positive stance towards a particular target, or vice versa. 
Further, for stance detection, the target need not be explicitly mentioned. These points are best illustrated via example: the tweet ``I hope that the Democrats get destroyed in this election!'' has a negative sentiment (towards Democrats), and (therefore, most likely) implies a positive stance towards Donald Trump.

As a case study for how context impacts annotations, we focus on political stance detection on Twitter---specifically, determining stance 
towards Hillary Clinton and Donald Trump during the 2016 U.S. election season. This task illustrates the challenges of annotation,
since individual tweets are often ambiguous with respect to stance, contexts on Twitter are inherently fractured, and differing contexts can make annotators lean in different directions.


Note that a user's stance, as we use the term in this paper, is a latent (and stable) property of the user. However, short of interviewing the user, we can never be completely certain of her stance. As such, the examples here and evaluations later rely on the authors' best estimates of stance, using all available information. 

A user's tweets, in turn, may or may not reveal her stance. This means that, by our definitions, an annotator might accurately perceive no stance in a tweet, yet have their annotation be considered incorrect with respect to the user's true stance. We would consider this case an annotator error caused by lack of context.

As examples of the task, consider annotating the following three tweets: (i) ``\emph{\small crooked Hillary - \#lockHerUp},'' (ii) ``\emph{\small Lester thinks he can control the crowd when he can't even keep Trump on topic lmao},'' and (iii) ``\emph{\small Settling in for \#debatenight Hoping to hear an adult conversation}.'' 
In the case of (i), a passing familiarity with American politics gives us high confidence that the author is pro-Trump.
The tweets in (ii) and (iii) are more ambiguous, but the authors' stances become clearer with access to varying forms of context. For (ii), a Pepe the frog
image (a symbol used by the American alt-right movement) in the user profile reveals that the user is probably a Trump supporter. Similarly, for (iii),
a profile description that reads ``\emph{\small Stereotypical Iowan who enjoys Hillary Clinton, progressive politics. Chair of \@CYDIWomen. Previously @HillaryForIA and @NARAL.}" suggests support for Clinton and distaste for Trump.


In order to explore the effects on annotation quality of providing these kinds of context to annotators, we crowd-source annotations for a  set of tweets and vary the additional information provided to annotators. For ease of comparison with related work and within our own study, we associate each user with a single anchor tweet. Thus, both annotators and classifiers are asked to determine the stance of a user using data from one particular time window.


\subsection{Data}

We collected tweets during the general election season (7/29/2016--11/7/2016) from over 40,000 Twitter users we had previously matched to voter registration records. Matching Twitter users to voter registrations \cite[using methods similar to][]{barbera_less_2016,hobbs_voters_2017} helps ensure that the accounts we study are controlled by humans, and it supplies additional demographic variables: gender, race and party registered with.


We identified as a political tweet any tweet that mentioned the official handle for Donald Trump (@realDonaldTrump) or Hillary Clinton (@HillaryClinton), or that contained one or more of the following terms or hashtags: Hillary, Clinton, Trump, Donald, \#maga,  \#imwithher, \#debatenight, \#election2016, \#electionnight. We removed all reply tweets, quote tweets and tweets that directly retweeted the candidates. Finally, we kept only those users who posted at least three political tweets.

From these users, we sampled 562 political tweets for crowd-sourced stance annotation, selecting at most one tweet per user. 
 These tweets were all sampled from users who were registered Democrats or Republicans. Half the tweets were paired with Hillary Clinton as the target, the other half with Donald Trump. 
 We also sampled and set aside an additional 250 + 318 tweet/target pairs to use as development and validation data, respectively (see Section~\ref{subsec:ground_truth}).

\subsection{Annotation Task}
We used Amazon Mechanical Turk (AMT) for annotation.  Annotators were presented a triplet of \{tweet, target, context\} and were asked to make their decisions on a 5-point Likert scale, ranging 
from  ``Definitely Opposes [target]'' to ``Definitely Supports [target]''. Both prior work \cite{mohammad_stance_2016} and our pilot studies suggested confusion between options for a tweet's irrelevance towards a target and the tweet's neutrality towards the target, so we used the center of the scale for both options. For this paper, we use a narrower three-point scale formed by merging the ``Definitely'' and ``Probably'' options. 
  
Further, while tweets were annotated with respect to different targets, we combine all annotations into a single task by assuming that ``anti-Trump'' means ``pro-Clinton'', and vice-versa. This assumption seems reasonable given that the voting population was strongly polarized during the general (post-primary) election season, and it doubles the amount of data we can use to train the models.
\ignore{This assumption ... \lisa{I was thinking of discussing corner cases we saw during labeling---it was mostly where the user disliked both candidates---but (a) maybe that belongs under ground truth, and (b) I don't recall our rules for resolving them.}}
Thus, throughout this work the labels we use are taken from the set \{``Support Trump / Oppose Clinton'' $= -1$, ``Neutral / I don't know'' $= 0$, ``Oppose Trump / Support Clinton'' $= 1$\}.



\subsection{Contexts Studied}

\begin{table}[!t]
  \centering
  \small
  \begin{tabular}{|p{2cm}|p{4.6cm}|}
  \hline
  \textbf{Context}	& 	\textbf{Displays the anchor tweet plus \ldots} \\
    \hline
   No Context& No additional information \\
   \hline
   Partial Profile & Profile image, name, and handle \\
   \hline
   \vspace{.01cm}Full Profile  & Author's profile image, name, handle, and description \\
   \hline
   Previous Tweets & Author's two most recent tweets \emph{in general} prior to the anchor \\
   \hline
   (Previous) Political Tweets &  Author's two most recent \emph{political} tweets prior to the anchor \\
   \hline
   \vspace{.01cm}Political Party & Political affiliation (if any) from the author's voter registration \\
   \hline
\end{tabular}
\caption{Descriptions of the six contexts (information conditions) presented to the annotators.}
  \label{tab:contexts}
\end{table}

Each of the 562 ``anchor'' tweets was annotated under six different \emph{contexts} (also referred to as information conditions) 
described in Table~\ref{tab:contexts}. (The Supplementary Material provides visual examples of each.) We collected at least three annotations for each tweet/condition pair. Every AMT worker was shown 40 different tweets, one by one, randomly distributed across contexts. 
Two additional artificial tweets were used to control for task competency.
\ignore{\lisa{We never mention how many workers are used total, so I don't see the value of stating how many were rejected.}}

We selected the conditions in Table~\ref{tab:contexts} based on two factors.  First, we included conditions that varied in how much we expected them to impact annotations. For example, we expected the partial profile information to have a relatively small effect, and political party a larger one. Second, we restricted our options to sets of information that we believed would minimally impact task completion times. We confirmed this empirically by regressing the (logged) time to completion for each annotator on the number of tweets she saw for each context, finding no significant effects from any context.

\subsection{Gold Standard Labels}
\label{subsec:ground_truth}
Ideally, we would evaluate annotation quality and downstream performance by comparing to ground truth.
Unfortunately, ground truth is difficult to characterize for tasks as subjective as stance detection or sentiment analysis \cite{passonneau_benefits_2014,dimaggio_adapting_2015}. 
In light of this, we constructed our own labels, using all available information about users, and we use them as an approximation of ground truth. 

We constructed these labels in order to evaluate downstream classification performance, and they cover a set of users not shown to the AMT workers. Given our resource constraints and the numerous (at least 18), often conflicting labels already available for tweets shown to AMT workers, we did not create definitive labels for that set. 


To create these ``gold standard'' (GS) labels, we considered all information found on the user's Twitter timeline, including everything AMT annotators could see, plus friend/following relationships, all of their previous tweets, demographics from the voter file, etc. 
Anecdotally, we found certain cases time-consuming to investigate, which argues for continuing to limit how much information we ask annotators to consider. All gold standard labels were agreed upon by at least two authors, who first labeled the data independently and then came together to discuss disagreements. 

Our GS set consists of 318 users (with their associated anchor tweets). Each user is assigned a label from the tertiary Trump/Neutral/Clinton scale. Another 250 manually labeled accounts were used for model development but are not part of reported results. The GS is approximately equally divided among registered Democrats, registered Republicans, and people not registered with either party; the last category includes self-declared Independents and voters not affiliated with any party. We include this third set in order to ensure the models generalize beyond registered Democrats and Republicans.

\section{Annotation Quality For Individual Contexts}  
\label{sec:indiv_contexts}	
In this section, we examine how annotator agreement varies depending on the context in which the labels were obtained, and how classifiers trained on majority-vote labels from each individual context, as well as on labels from all contexts combined, perform on the GS. First, we introduce the classifier and features used for the latter task, then discuss results for agreement and classifier performance.

\begin{table*}
\small
\centering
\begin{tabular}{|p{1.2cm}|p{3.35cm}|p{10.cm}|}
\hline
\textbf{Category} & \textbf{Data Source} & \textbf{Feature Representation} \\ \hline
Text & Anchor tweet, \newline previous (political) tweets, \newline profile description & 
Character n-grams ($n \in [3,5]$), word n-grams ($n \in [1,3]$). \newline
Preprocessing: only use tokens appearing $\ge 10$ times, apply tf-idf weighting.
  \\ \hline
\multirow{2}{3cm}{Sentiment} & \multirow{2}{3cm}{Anchor tweet} & 
VADER score \cite{hutto_vader:_2014-1} \\ \cline{3-3}
& & Dictionary approach \cite{joseph_girls_2017}: valence, dominance \& arousal scores \\ \hline
User & Voter registration record & Race, gender \\ \hline
\end{tabular}
\caption{Features used in classification.}
\label{tab:features}
\end{table*}

\subsection{Classifier, Labels, Features, \& Evaluation}
For each of the six contexts separately, we construct labels with which to train a classifier. Training labels are constructed using majority vote; we also tried weighting the training instances to match the distribution of labels, but it did not perform as well. We also construct a seventh set of labels using all annotations from all conditions. We then train a classifier on each set of labels. We use Random Forest models, as they outperformed regularized logistic regression and SVMs with linear kernels on the development set. Note that the only difference among the models in this section is the \emph{labels} they are \emph{trained} on.

The feature set used, shown in Table~\ref{tab:features}, is meant as a straightforward representation of the information seen by annotators; parts of it follow \citet{ebrahimi_joint_2016}. We construct three types of features for each tweet: text, sentiment and user features.
For text features, we collapse the anchor tweet plus all additional textual context seen by any annotator into a single string, then compute various n-grams from it. For sentiment, we compute various scores from the anchor tweet alone. For user features, we include the user's race and gender, which annotators might have learned from the user's profile picture. Note that because we want models to generalize beyond registered Democrats or Republicans, we \emph{do not} include a feature for political party. 

Classifier performance on the GS is measured, following prior work \cite{mohammad_semeval-2016_2016,ebrahimi_joint_2016}, on the average of the F1 scores on the two classes of interest (``Clinton'' and ``Trump''). Additionally, we report the average log-loss (the negative log-likelihood, according to the classifier, of the true label). Log-loss and F1 can be seen as complementary measures: whereas F1 evaluates the quality of the ranking of test instances, log-loss evaluates the quality of their individual probability estimates. To compute the probability estimate from a Random Forest, we compute mean class probabilities across all trees. 

To assess the statistical significance of differences between two models, we first obtain probability estimates for all GS items. For log-loss, we use a Mann-Whitney test on the scores from the two models being compared. For F1, we create 1000 bootstrap iterations of the sample, compute the average F1 of each, and run a non-parametric difference-in-means test, using 95\% confidence intervals.

\begin{table}
\small
\centering
\begin{tabular}{|c|c|c|c|c|}
\hline
\textbf{Model} & \textbf{Agreement} & \textbf{Log-Loss} & \textbf{Avg F1} \\ 
\hline 
No Context  & 0.84 & 0.72 & 0.61 \\ 
\hline
Partial Profile & 0.83 &  0.71 &	0.68 \\ 
\hline
Full Profile & 0.82 &  0.69 & 0.62 \\ 
\hline
Previous Tweets & 0.84 & 0.65 &	\textbf{0.71} \\ 
\hline
Political Tweets & \textbf{0.88}  & \textbf{0.61} &	0.70	 \\ 
\hline
Political Party & \textbf{0.88} & 0.63	& 0.68 \\ 
\hline
All Combined  & 0.77 & 0.62 & \textbf{0.71} \\ 
\hline
\end{tabular}
\caption{Inter-annotator agreement, then performance of classifier trained on majority vote labels. (Best possible is 1 for agreement and F1, 0 for log-loss.)}
\label{tab:res}
\end{table}

\subsection{Effects of the Contexts}

Before evaluating classification results, we consider annotator agreement within each context, calculated like \citet{mohammad_stance_2016} as the average, across tweets, of the percentage of annotations that match the majority vote.
As shown in Table~\ref{tab:res}, annotators shown No Context achieve an agreement score of 0.84, similar to the 0.8185 reported by \citet{mohammad_stance_2016}. Relative to this baseline, some contexts increase agreement more than others. As one might expect, Previous Political Tweets and Political Party show the strongest signals. Their effects are statistically ($p < .01$, Mann-Whitney test) and practically significant, increasing the number of labels having \emph{full} agreement by 15\% and 10\%, respectively. 


However, annotators shown different contexts did not necessarily converge to the same labels. Notice the low agreement for the All Combined condition: the majority labels held stronger majorities within any individual context than across all of them.
In fact, if we look at the six majority vote labels for each tweet, only in 43\% of the tweets are these labels in full agreement. 
At the end of Section~\ref{sec:results}, we return to the question of why agreement was so low across conditions, with the help of parameters estimated by ConStance.

In the classification task, the results in Table~\ref{tab:res} further suggest that Previous Political Tweets serves as the strongest single context.  
There is a good case to be made for choosing this individual context, which is statistically significantly better than many others.
For example, providing annotators with Previous Political Tweets provides a statistically significant increase in both average F1 scores and log-loss (with $p < .01$) over both the No Context and Full Profile conditions. Perhaps most noteworthy is that the All Combined classifier, created from the naive combination of all annotations, is no better than the classifiers from the individual conditions.



To summarize, results suggest that providing annotators with appropriate additional context can improve annotation quality, as measured via annotator agreement and downstream classification performance. However, it was not obvious in advance which context would be most helpful, and performing such an analysis as this requires the time-intensive construction of better ``gold standard'' labels against which to check the labels already being outsourced to annotators. In addition, the heterogeneity of the labels produced in different contexts suggests that the contexts provide diverse signals we might be able to leverage; however, simply combining all the annotations does not result in improvements.



\section{ConStance: General Unified Model}
\label{sec:model}

The prior section thus suggests that it may be better to limit a priori decisions and instead to leverage multiple kinds of context during annotation. Like \citeapos{raykar_learning_2010} assumes for annotators, we might expect (and indeed find) that even those contexts that turn out to be worse on some metrics still might be useful for other purposes. Here, we present a model for such an approach.

ConStance learns a classifier for \emph{items}. For our purposes here, an item is a user together with their anchor tweet and the additional information from which features were derived (see Table~\ref{tab:features}); more broadly, it is whatever we choose to put into the feature vector. One could choose a different setup; for example, an item could be a user and ten anchor tweets. However, the current arrangement allows for straightforward comparison to prior stance work on Twitter \cite{mohammad_semeval-2016_2016}. 

Note that in general, the features need not be restricted to those annotators could have seen. Rather, they could include anything useful to a classifier. Note also that the feature set provided to ConStance is the same used by the baseline models; only the models themselves differ.

\subsection{Overview}

\begin{figure}[t]
	\centering
	\includegraphics[width=.45\textwidth]{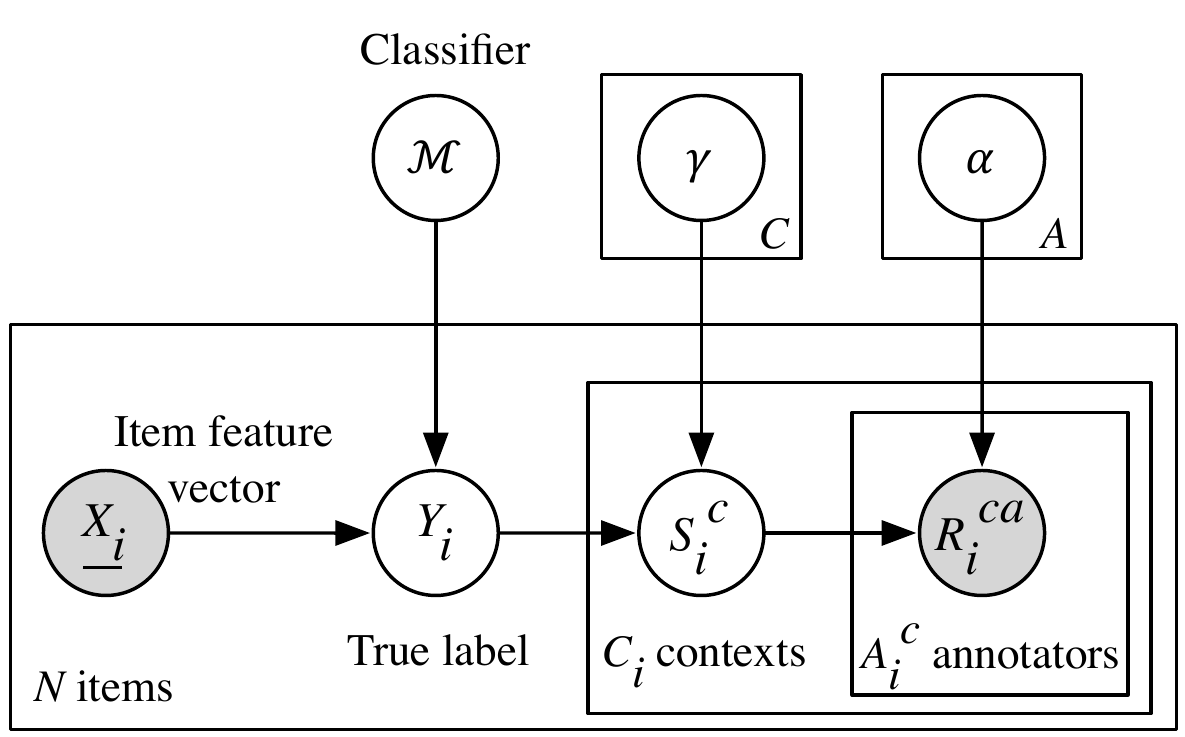}
	\caption{Graphical model for ConStance.}
	\label{fig:mod}
\end{figure}

\begin{table}
	\small
	\begin{tabular}{|p{.47cm}|p{6.2cm}|} \hline
	\textbf{Var.} & \textbf{Meaning} \\ \hline
	$\underline{X_i}$ & Feature vector of item $i$ \\ \hline
	$Y_i$ & Latent true label of item $i$ \\ \hline
	$S_i^c$ & Latent context-specific label of item $i$ after noise from context $c$  \\ \hline
	$R_i^{ca}$ &  Label given by annotator $a$ to item $i$ in context $c$ \\ \Xhline{1.2pt}
	$V$ & Set of values for labels and annotations: $\{-1, 0, 1\}$ \\ \hline
	$N$ & \# of items, indexed by $i$ \\ \hline
	$C$ & Set of contexts, indexed by $c$ \\ \hline
	$A$ & Set of annotators, indexed by $a$ \\ \Xhline{1.2pt}
	$\mathcal{M}$ & Learned classifier \\ \hline
	$\gamma^c$ & $V \times V$ parameter matrix for context $c$ \\ \hline
	$\alpha^a$ & $V \times V$ parameter matrix for annotator $a$ \\ \Xhline{1.2pt}
	$\mathcal{D}$ & All observed data: all values of $\underline{X_i}$ and $R_i^{ca}$ \\ \hline
	$Z$ & All latent variables: all values of $Y_i$ and $\underline{S_i}$ \\ \hline
	$\theta$ & All model parameters: $\mathcal{M},\gamma,\alpha$ \\ \Xhline{1.2pt}
	$T_i$ & 	All latent variables for item $i$: $(Y_i, \underline{S_i})$ \\ \hline
	$\tau_{i(y\underline{s})}$ & Current estimate of all latent values for item $i$: $p(Y_i = y, \underline{S_i} = \underline{s} \mid \mathcal{D}, \theta)$  \\  \hline
	\end{tabular}
	\caption{Model variables.}
	\label{tab:mod}
\end{table}

The model we develop is shown in Figure~\ref{fig:mod}. There are $N$ items to be labeled. Each item can be viewed in up to $C$ different contexts. Finally, there are $A$ total annotators labeling the items; each annotator sees multiple items.  Each item can have a different number of annotations, produced by any assignment of annotators and information conditions to items. In our dataset, every item is labeled in 6 conditions (every $|C_i| = 6$), and within every context, every item is labeled by at least 3 annotators (every $|A_i^c| \ge 3$).



The model's generative story works as follows. Item $i$ has feature vector $\underline{X_i}$ and a ``true'' label $Y_i \in V$. 
The relationship between $\underline{X_i}$ and $Y_i$ can be described by some model $\mathcal{M}$, which we will ultimately learn. When the item is viewed with context $c$, the item's true label $Y_i$ is transformed by noise into a ``context-specific'' label $S_i^c \in V$. In other words, the true label may appear differently when seen through the lens of each context. The variable $S_i^c$ represents what an ideal annotator would say about item $i$ given only as much information as is preserved by context $c$. 

The ``noise'' introduced by context $c$ is described by parameter $\gamma^c$. The parameter $\gamma^c$  is a $V \times V$ matrix of transition probabilities from true labels to context-specific labels. These probabilities depend only on $Y_i$ and $\gamma^c$, not on the item's features $\underline{X_i}$.

Importantly, annotators themselves are also imperfect. When annotator $a$ sees item $i$, she may also distort the label she sees, $S_i^c$, into the observed annotation $R_i^{ca} \in V$. The annotator-specific noise process is described by parameter $\alpha^a$, another $V \times V$ transition matrix.

For a better understanding of the role of $\gamma^c$ (and by anology, $\alpha^a$), consider the depictions in Figure~\ref{fig:gamma}. 
The matrix on the top left refers to the No Context condition. 
Its top row describes what an annotator with perfect judgment would think about a user whose true label is Trump [supporter], with no context. The top left cell, with a value around 0.65, is the probability the annotator would think Trump; the lighter middle cell, with a value around 0.35, is the probability she would think Neutral/Don't know; and the probability she would think Clinton is almost 0.



\subsection{Learning}

Like \citeapos{raykar_learning_2010}, we perform inference using Expectation Maximization (EM). A full derivation is provided in the Supplementary Material; here, we sketch the main steps.

The model's incomplete data likelihood function, Eq.~\eqref{eq:incompleteLik}, describes the joint probability, across all items, of $Y_i$, all values of $S_i^c$, and all values of $R_i^{ca}$ assuming $\underline{X_i}$ is known and fixed. Uppercase denotes random variables; lowercase, specific values. 
In line \eqref{eq:incompleteLik2}, we substitute in the equivalent model parameters. 
\begin{align}
	p( \mathcal{D} | & \theta, X)=  \prod_{i=1}^N \sum_{y}^V p(Y_i = y | \underline{x_i}, \mathcal{M}) \prod_{c}^{C_i}  \notag \\ 
	  & \qquad \sum_{s}^V p(S_i^c = s | y, \gamma) \prod_{a}^{A_i^c} p(r_i^{ca} | s, \alpha) 	\label{eq:incompleteLik} \\
& \qquad = \prod_{i=1}^N \sum_{y}^V \mathcal{M}_y(\underline{x_i}) \prod_{c}^{C_i} \sum_{s}^V \gamma^c_{ys} \prod_{a}^{A_i^c} \alpha^a_{sr} \label{eq:incompleteLik2}
\end{align}

The EM derivation is difficult because both $Y_i$ and $\underline{S_i}$ are unobserved. Our solution is to treat the latent variables as a block, describing their joint configuration with a single term $T_i = (Y_i, \underline{S_i})$. In our data, since $|C_i| = 6$, $T_i$ can take on $7^{|V|}$ possible values, a number small enough to enumerate over when we need to marginalize out $T_i$. 

We define membership indicator variables $T_{i(y\underline{s})} \in \{0, 1\}$ such that $T_{i(y\underline{s})} = 1$ if $T_i$ has the specific values $(y, \underline{s})$. During learning, we use analogous variables $\tau_{i(y\underline{s})} \in [0,1]$ to represent the posterior probabilities of each configuration: $\tau_{i(y\underline{s})} = p(T_{i(y\underline{s})} = 1 \mid \mathcal{D}, \theta)$. 
The expected value of the complete data log-likelihood is:
\begin{align}
\mathbb{E}&_Z[\ell(\mathcal{D}, Z | \theta, X)] 
	 = \sum_{i=1}^N \sum_{y}^V \left( \sum_{s_i^1}^V \ldots \sum_{s_i^{C_i}}^V  \right) \notag \\
	 &\tau_{i(y\underline{s})} 
	 ( 
		\log p(T_{i(y\underline{s})} \mid \underline{x_i}, \mathcal{M}, \gamma) + 
		\sum_{c}^{C_i} \sum_{a}^{A_i^c} 
	\log \alpha^a_{sr} 
	 ) \label{eq:expCompleteLogLik}
\end{align}

For the E step, we update the membership estimates $\tau_{i(y\underline{s})}$ using the current parameters $\theta$. With Bayes' rule, each item's new value of $\tau_{i(y\underline{s})}$ is shown to be the full joint likelihood of item $i$ (see Eq.~\eqref{eq:incompleteLik2}) when setting $Y_i = y$ and $\underline{S_i} = \underline{s}$, divided by the sum, over all possible settings of $Y_i$ and $\underline{S_i}$, of that same joint likelihood.

For the M step, we update the model parameters using the current membership estimates. To update the classifier $\mathcal{M}$, following the guidance of \citeapos{raykar_learning_2010}, we retrain the classifier using the current estimates of $Y_i$ as weights for items. The estimates of $Y_i$ can be obtained from $\tau_{iy\underline{s}}$ by marginalizing out $\underline{S_i}$, thus $\mathbb{E}_Z[Y_{i} = y] =  \sum_{s_i^1}^V \ldots \sum_{s_i^{C_i}}^V \tau_{iy\underline{s}}$. We then use sampling to construct a discrete set of labels for model training based on these weights.

To update $\gamma$ and $\alpha$, we maximize them with respect to Eq.~\eqref{eq:expCompleteLogLik}. For $\gamma$, the entry $\gamma^c_{ys}$ (i.e., row $y$, column $s$ of matrix $\gamma^c$) denotes $p(S_i^c = s \mid Y_i = y)$. Each matrix entry can be updated individually by taking the partial derivative of Eq.~\eqref{eq:expCompleteLogLik} and using, as a Lagrange multiplier term, the constraint that the row must sum to 1. The updated value for $\gamma^c_{ys}$ turns out to be a fraction in which the numerator is the weighted (by $\tau$) number of items having $Y_i = y$ and $S_i^c = s$, and the denominator is the weighted number of items having $Y_i = y$ (and any value for $S_i^c$). For $\alpha$, a similar derivation yields the following update to $\alpha^a_{sr}$: the weighted number of labels by annotator $a$, in any context, having $S_i^c = s$ and $R_i^{ca} = r$, divided by the weighted number of labels by annotator $a$, in any context, having $S_i^c = s$.

\section{Model Results and Discussion}
\label{sec:results}

\begin{table}
\small
\centering
\begin{tabular}{|p{3.7cm}|p{1.3cm}|p{1.3cm}|}
\hline
\textbf{Model}  & \textbf{Log-Loss} & \textbf{Avg F1} \\ \hline 
Best baselines  & \textit{0.61} & \textit{0.71} \\ \hline \hline
\textbf{ConStance} & \textbf{0.57} & \textbf{0.77} \\ \hline \hline
 \multicolumn{3}{|l|}{Ablations} \\ \hline
1. Only Political Tweets	 & 0.59 & \textit{0.73} \\ \hline
2. Context Labels Masked & \textbf{0.57} & 0.75 \\ \hline
3. Annotator Labels Masked &  \textit{0.65} & 0.75 \\ \hline
\end{tabular}
\caption{Classification performance of ConStance and model ablations. Boldface highlights best scores. Significance tests use the the $p < .05$ level for log-loss. Compared to the best baselines, all scores that appear better are statistically significant. Italics indicate the scores that are significantly worse than ConStance.}
\label{tab:modelRes}
\vspace{-1.2em}
\end{table}

\ignore{Possibly move the line about runtime here, and add values of hyperparameters.}

The top portion of Table~\ref{tab:modelRes} displays ConStance's performance compared to the best results from Section~\ref{sec:indiv_contexts}. Using the same experimental setup as Section~\ref{sec:indiv_contexts}---the model type and features, $\mathcal{M}$ and $X$ respectively, are the same as in the baselines---ConStance improves over the best baseline models for each metric. This improvement is statistically significant for both metrics (at the $p < .05$ level for log-loss). Further, the model converges rapidly, within 5-7 iterations of the EM algorithm and 3-5 minutes on a single machine.\footnote{As above, a development set is used for coarse hyperparameter tuning; see the Supplementary Material for details.}


In addition to comparing to the baselines provided in Section~\ref{sec:indiv_contexts}, we investigate which information the model is leveraging to be successful. We do so by exploring three ablations of the model. Variation \#1 (``Only Political Tweets'' in Table \ref{tab:modelRes}) uses the full model, but only gives it the annotations from the Political Tweets condition. This tests whether simply modeling differences in annotators' error rates, as \citeapos{raykar_learning_2010} do, with a single (``best'') context is helpful. We find that it is: the performance of this variation is significantly better on both metrics than the Political Tweets baseline from Table~\ref{tab:res}. 

In the second and third variations, we check whether the effectiveness of ConStance stems from modeling differences between annotators rather than differences in contexts, or vice versa. Variation \#2 (``Context Labels Masked''), like \#1, models only annotator effects; however, it instead uses the entire set of annotations, treating them as if from a single context (i.e., ``masking'' context information from the model). Variation \#3 (``Annotator Labels Masked'') is the complement of Variation \#2: it models differences in contexts, and it uses the entire set of annotations, treating them as if from a single annotator. 

The results of the model ablation experiments are three-fold. First, we see that each piece of the model on its own is effective in moving beyond baseline approaches that use only one context or naively combine labels across contexts and annotators (the ``All Combined'' baseline). All model variations achieve significantly higher Avg.~F1 than the baselines, and Variations \#1 and \#2 improve on log-loss. Second, we see that modeling annotators alone is clearly better than not: not only does Variation \#1 outperform the Political Tweets baseline (significantly), but also Variation \#2 outperforms the All Combined baseline (significantly) and ConStance outperforms Variation \#3 (with significance in one measure). Finally, the best results come from using the full model. Even if the differences between ConStance and the variations are not all statistically significant, modeling both annotators and contexts appears to be the most complete and effective approach.


\begin{figure}[t]
\centering
\includegraphics[width=.48\textwidth]{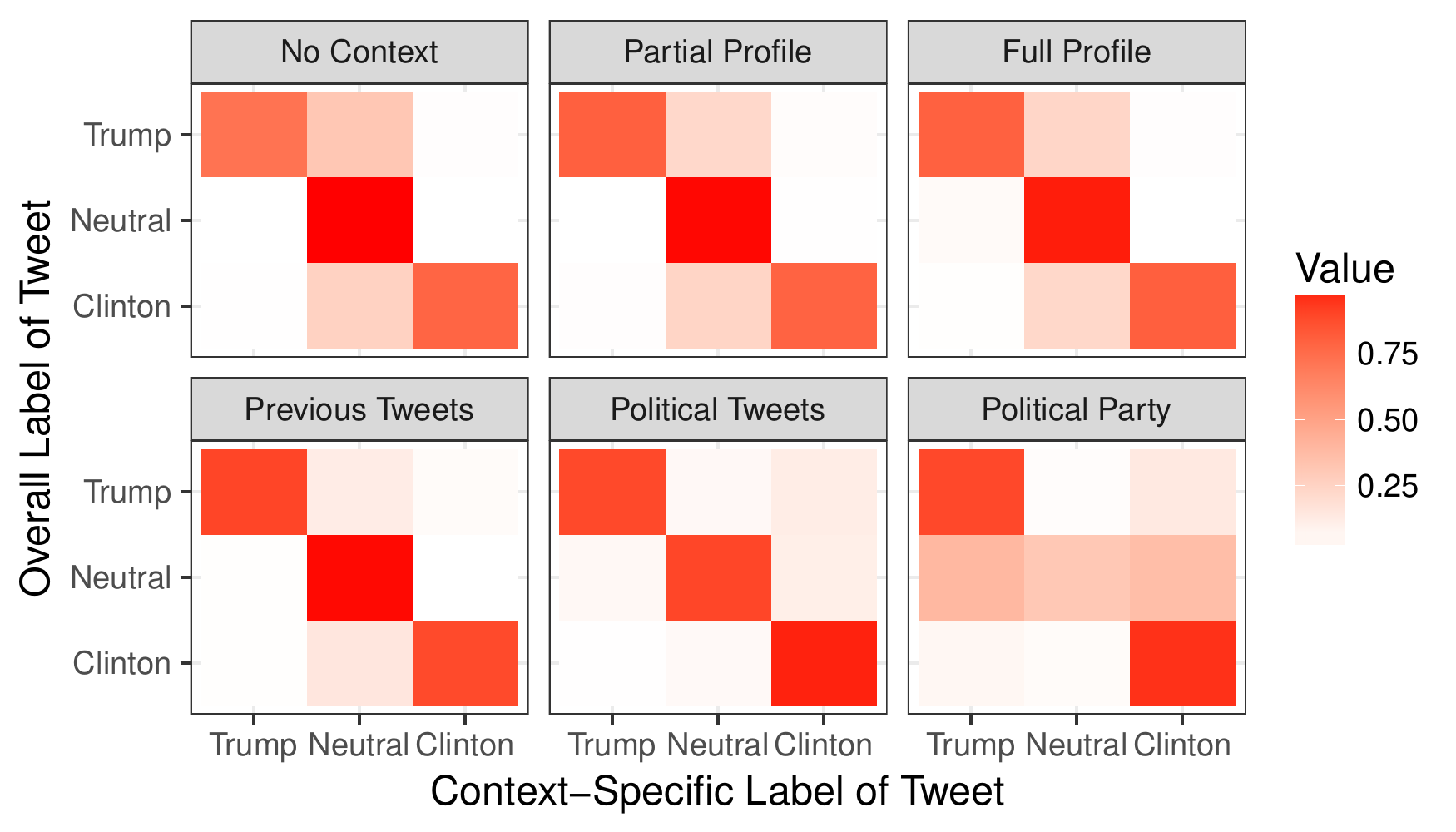}
\caption{Parameter matrices $\gamma^c$ learned by ConStance for each context. Darker shading indicates higher values.}
\label{fig:gamma}
\vspace{-1em}
\end{figure}

In addition to model performance, we can also examine what ConStance has learned about the quality of labels from each context. Recall that the model produces a parameter matrix for each context, $\gamma^c$, which describes how a context distorts the ``true'' labels the model assumes. Each $\gamma^c$ is a transition matrix, so a context that perfectly preserves true labels would show up as the identity matrix; off-diagonal entries show error patterns. 

Figure~\ref{fig:gamma} visualizes parameter estimates for $\gamma$. We see that in the No Context, Partial Profile and Full Profile conditions, annotators often selected the ``Neutral'' option ($x$-axis) when the model inferred the true label was ``Clinton'' or ``Trump'' ($y$-axis).  This finding is in line with intuitions; annotators who saw these conditions simply lacked enough information to determine any label. 

On the other extreme, in the Political Party context, annotators selected ``Trump'' or ``Clinton'' too often when the model settled on the ``Neutral'' option. That is, even when a user's stance was not clear to annotators in other conditions, annotators who saw political party still inferred stance from the text. Here, one could argue annotators were shown ``too much'' or ``too strong'' a context---they saw stance even where the content produced by the user did not suggest one. Indeed, further manual inspection of 90 tweets on which annotations disagreed across contexts implies that annotators who saw political affiliation were often wrong because they focused too little on text content relative to the provided political affiliations.



In presenting these findings, a key point to highlight is that unlike the results of  Section~\ref{sec:indiv_contexts},  Figure~\ref{fig:gamma} was produced without access to any full information labels, which depend on a significant level of manual effort beyond annotations gathered on AMT. 

\ignore{\lisa{I decided the following is basically conclusion material. Not sure we need to say anything more to wrap up this seciton.}
Discussion: using ConStance was way easier and better than choosing a single context. This is the way of the future because...(because choosing a single ground truth label is throwing away info. It's better to do things jointly if inference is tractable, which it is and will continue to be, provided...) And you know what else we could do? (We could talk about other things that would be great to do!)}

\section{Related Work}
\label{sec:related}
Recent work has shown that cognitive biases such as stereotypes \citep{carpenter_real_2016} and anchoring \cite{berzak_anchoring_2016} can negatively impact text annotation and resulting models, even for objective tasks like POS tagging \cite{blodgett_demographic_2016}. Still, researchers often decide what context to show annotators without rigorously evaluating how their decisions will affect annotations, on tasks from gender identification to political leanings \cite{chen_comparative_2015,nguyen_why_2014,burger_discriminating_2011,cohen_classifying_2013-1}. Our work suggests an interesting avenue of development towards reducing annotation bias by explicitly modeling it and reducing the need for a priori decisions on which context is best for which particular task.

In doing so, we draw on a large body of work around improving annotation quality for NLP data. Our work aligns with efforts to improve task design \cite[e.g.][]{schneider_framework_2013,morstatter2016replacing,schneider_what_2015}, and to develop better models of annotation. With respect to the former and specific to Twitter, \citet{frankenstein_contextualized_2016} show that for the task of labeling the sentiment of reply tweets, annotations vary depending on whether or not the original tweet (being replied to) is also shown.  With respect to the latter, several recent models beyond \citet{raykar_learning_2010} have been proposed \cite{guan_who_2017,tian_learning_2012,wauthier_bayesian_2011,passonneau_benefits_2014}. However, our work is most similar to efforts outside the domain of NLP, where \citet{dai_pomdp-based_2013} have developed a method of switching between task workflows based on annotation quality for particular items, and \citet{nguyen_probabilistic_2016-1} have developed a Bayesian model similar to ours to study annotation quality for other kinds of slightly subjective tasks.

In a closely related vein, recent work has also considered how text annotations may vary in important ways based on the characteristics of annotators (rather than how the task is posed, as we study here) \cite{sen_towards_2015}.  An interesting avenue of future work is to understand the intersection between the design of NLP annotation tasks and the characteristics of the annotating population.


\section{Conclusion and Future Work}
\label{sec:future}
Annotated data serves as a foundational layer for many NLP tasks. While some annotation tasks only require information from short texts, in many others, we can elicit higher-quality labels by providing annotators with additional contextual information. However, asking annotators to consider too much information would make their task slow and burdensome.

In this paper we demonstrate how exposing annotators to short contextual information leads to better labels and better classification results. However, different contexts lead to results of different quality, and it is not obvious a priori which context is best, nor---even given ground truth---how to combine labels produced across contexts to exploit the information present in each. We then propose ConStance, a generalizable model that learns the effects of both individual contexts and individual annotators on the labeling process. The model infers (probability estimates for) ground truth labels, plus learns a classifier that can be applied to new instances. We show that this classifier significantly improves classification of political stance compared to the standard practice of training models on majority vote labels.

The focus of this work is on improving both the annotation process for nuanced, context-dependent tasks and the use of the resulting labels. While ConStance's label estimation can be used in conjunction with any classification method, this paper does not address the optimization of the classifier itself. 
Thus, while we consider an assortment of contexts and use a rich feature representation, using additional contexts or different features may lead to better performance on stance detection. Finally, the model is versatile enough we could consider treating different tweets as different ``contexts'' for the same user, augmenting the extensively annotated tweets with other types of data, and, naturally, applying the same framework to other annotation tasks.



\bibliography{emnlp2017}
\bibliographystyle{emnlp_natbib}

\appendix
\onecolumn

This appendix provides more details on the annotation study we developed (Appendix~\ref{sec:study}), of the EM algorithm we briefly cover in the main text (Appendix~\ref{sec:em}), a brief description of how we debugged this algorithm (Appendix~\ref{sec:deb}) and a description of how hyperparameters were set using a test set (Appendix~\ref{sec:hyp}).

\section{More Details on Annotation Study}\label{sec:study}

\begin{figure}[t]
\centering
\includegraphics[width=\textwidth]{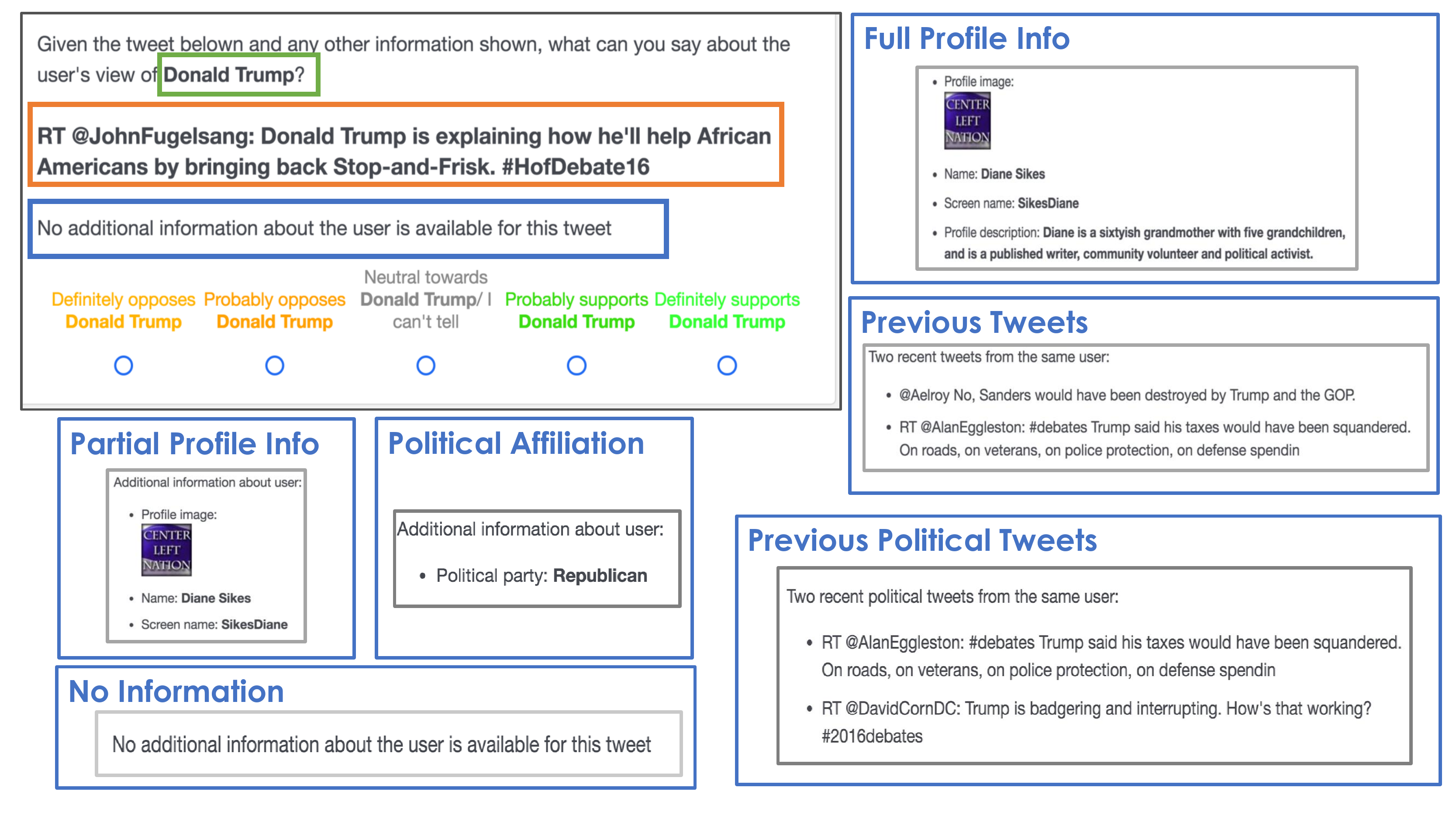}\\
\caption{In the black box is a single annotation question. The green box displays where the target is given, the orange box where the tweet 
text is displayed and the blue box where any additional information is given. The six information conditions are shown in the blue boxes.}
\vspace{-1em}
\label{fig:overview}	
\end{figure}

Figure~\ref{fig:overview} presents an overview of our study design. Outlined in black on the top left is an example of one of the questions posed to annotators.  Each question developed consists of three main parts: a \emph{target}, the text of a particular \emph{tweet}, and a set of \emph{additional information} about the tweet's author. At a high level, we first selected a set of tweet/target pairs. We then produced six questions for each tweet/target pair, one for each type of additional information/context we considered.

We initially selected a set of 480 tweet/target pairs to annotate, split evenly between the two targets.  The 240 tweets for each target were selected by choosing 40 tweets from each possible combination of these two tweet-level properties (2  ``tweet originality'' types x 3 ``target mention'' types). After an initial investigation of results, we observed that the sample contained relatively few tweets from Republican users. To address this issue we sampled an additional 82 tweets from Republican users, for a final sample size of 562 tweets. These tweets imbalanced the sample design---they were no longer evenly distributed across the original categories---but they ensured sufficient counts for Republican users, which we believed might be useful for our analyses.

As a final point, we replace all URLs in both the tweets to be labeled and tweets shown in the additional information portion of the questions with the text ``\{\{link\}\}''. This decision was made in order to maintain control over the amount of information seen by annotators; if the links were left visible, annotators would vary in whether or not they clicked through. However, since URL information (e.g. domain name, page title, page content) provides useful information to the annotator, obscuring the links artificially increased the task's difficulty.

\section{Derivation of EM Algorithm}\label{sec:em}

\begin{figure}
\begin{floatrow}
\ffigbox{%
	\includegraphics[width=.4\textwidth]{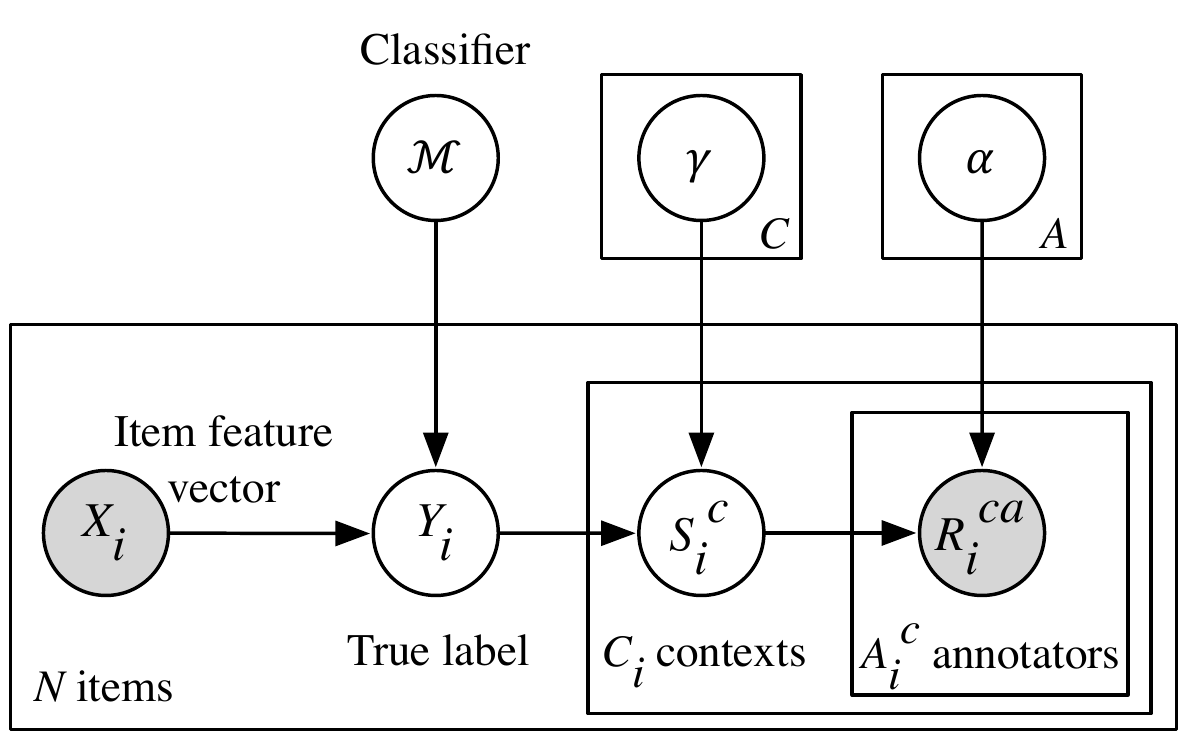}
}{%
  	\caption{Graphical model for ConStance.}
  	\label{fig:mod}
}
\capbtabbox{%
	\small
 	  \begin{tabular}{|p{.47cm}|p{8cm}|} \hline
	\textbf{Var.} & \textbf{Meaning} \\ \hline
	$\underline{X_i}$ & Feature vector of item $i$ \\ \hline
	$Y_i$ & Latent true label of item $i$ \\ \hline
	$S_i^c$ & Latent context-specific label of item $i$ after noise from context $c$  \\ \hline
	$R_i^{ca}$ &  Label given by annotator $a$ to item $i$ in context $c$ \\ \Xhline{1.2pt}
	$V$ & Set of values for labels and annotations: $\{-1, 0, 1\}$ \\ \hline
	$N$ & \# of items, indexed by $i$ \\ \hline
	$C$ & Set of contexts, indexed by $c$ \\ \hline
	$A$ & Set of annotators, indexed by $a$ \\ \Xhline{1.2pt}
	$\mathcal{M}$ & Learned classifier \\ \hline
	$\gamma^c$ & $V \times V$ parameter matrix for context $c$ \\ \hline
	$\alpha^a$ & $V \times V$ parameter matrix for annotator $a$ \\ \Xhline{1.2pt}
	$\mathcal{D}$ & All observed data: all values of $\underline{X_i}$ and $R_i^{ca}$ \\ \hline
	$Z$ & All latent variables: all values of $Y_i$ and $\underline{S_i}$ \\ \hline
	$\theta$ & All model parameters: $\mathcal{M},\gamma,\alpha$ \\ \Xhline{1.2pt}
	$T_i$ & 	All latent variables for item $i$: $(Y_i, \underline{S_i})$ \\ \hline
	$\tau_{i(y\underline{s})}$ & Current estimate of all latent values for item $i$: $p(Y_i = y, \underline{S_i} = \underline{s} \mid \mathcal{D}, \theta)$  \\  \hline
	\end{tabular}}{%
  \caption{Model variables.}%
  \label{tab:vars}
}
\end{floatrow}
\end{figure}

Figure~\ref{fig:mod} provides a graphical overview of the model; both it and Table~\ref{tab:vars} are reproduced from the main paper for convenience. Below, we outline the derivation of the EM algorithm used for inference.

The model's incomplete data likelihood function, Eq.~\eqref{eq:incompleteLik}, describes the joint probability, across all items, of $Y_i$, all values of $S_i^c$, and all values of $R_i^{ca}$ assuming $\underline{X_i}$ is known and fixed. Uppercase denotes random variables; lowercase, specific values. 
In line \eqref{eq:incompleteLik2}, we substitute in the equivalent model parameters.
\begin{align}
	p( \mathcal{D} | & \theta, X)=  \prod_{i=1}^N \sum_{y}^V p(Y_i = y | \underline{x_i}, \mathcal{M}) \prod_{c}^{C_i}  \notag \\ 
	  & \qquad \sum_{s}^V p(S_i^c = s | y, \gamma) \prod_{a}^{A_i^c} p(r_i^{ca} | s, \alpha) 	\label{eq:incompleteLik} \\
	 &= \prod_{i=1}^N \sum_{y}^V \mathcal{M}_y(\underline{x_i}) \prod_{c}^{C_i} \sum_{s}^V \gamma^c_{ys} \prod_{a}^{A_i^c} \alpha^a_{sr} \label{eq:incompleteLik2}
\end{align}

To derive EM for this model, we treat the latent variables as a block, moving their joint distribution into a single term. A given tweet has a latent value $y_i$ and a latent vector $\underline{s_i}$ containing one entry per context: $\underline{s_i} = (s_i^1, \ldots, s_i^{C_i})$. Returning to \eqref{eq:incompleteLik}, we move the term $p(S_i^c = s | y, \gamma)$ left, outside the product over $C_i$ contexts, explicitly representing each component $s_i^c$ of $\underline{s_i}$ and summing over its latent values. With all of $\underline{s_i}$ in scope at once, we can rearrange the latent variables into a single term.
\begin{align*}
	p(\mathcal{D} | \theta) &= \prod_{i=1}^N \sum_{y}^V p(y_i = y | \underline{x_i}, \mathcal{M}) \left( \sum_{s_i^1}^V \ldots \sum_{s_i^{C_i}}^V \right) \prod_{c}^{C_i} p(s_i^c = s | y_i, \gamma)  \prod_{a}^{A_i^c} p(r_i^{ca} | s_i^c, \alpha)  \\
	 &= \prod_{i=1}^N \sum_{y}^V \left( \sum_{s_i^1}^V \ldots \sum_{s_i^{C_i}}^V \right) p(y_i = y | \underline{x_i}, \mathcal{M})  p(\underline{s_i} = \underline{s} | y_i, \gamma) \prod_{c}^{C_i} \prod_{a}^{A_i^c} p(r_i^{ca} | s_i^c, \alpha) \\
	 &= \prod_{i=1}^N \sum_{y}^V \left( \sum_{s_i^1}^V \ldots \sum_{s_i^{C_i}}^V \right) p(y_i = y, \underline{s_i} = \underline{s} \mid \underline{x_i}, \mathcal{M}, \gamma) \prod_{c}^{C_i} \prod_{a}^{A_i^c} p(r_i^{ca} | s_i^c, \alpha) 
\end{align*}

Next we introduce an indicator variable $T_{i(y\underline{s})} \in \{0, 1\}$ representing a configuration of latent variable assignments $(y_i, \underline{s_i}) \in Z$. 
We define $T_{i(y\underline{s})} = 1$ when tweet $i$ has the specific configuration $(y_i = y, \underline{s_i} = \underline{s})$. This gives:
\begin{align*}
p(\mathcal{D} | \theta) &= \prod_{i=1}^N \sum_{y}^V \left( \sum_{s_i^1}^V \ldots \sum_{s_i^{C_i}}^V \right) p(T_{i(y\underline{s})} \mid \underline{x_i}, \mathcal{M}, \gamma) \prod_{c}^{C_i} \prod_{a}^{A_i^c} p(r_i^{ca} | s_i^c, \alpha) \\
\end{align*}
During the E step, we will use analogous variables $\tau_{i(y\underline{s})} \in [0,1]$ to represent the conditional probabilities of $T_{i(y\underline{s})}$.

Below, we derive the E-step and the M-step. For clarity, we first express the complete-data likelihood function (and the complete data log-likelihood) and the expected complete log-likelihood, which we then use to determine solutions for the E-step and the M-step.

\subsection*{Complete data likelihood function}

Here, we assume that we have the observed values of every $T_{i(y\underline{s})}$. The $T_{i(y\underline{s})}$ in the exponent is an observed 0 or 1, while the $p(T_{i(y\underline{s})} = 1 \mid \ldots)$ is still a prior probability to compute. (That prior does use its parent variables $\underline{x_i}$, but importantly, doesn't use $r_i^{ca}$.)
\begin{align*}
p(\mathcal{D}, Z \mid \theta) &=
	 \prod_{i=1}^N \prod_{y}^V \left( \prod_{s_i^1=1}^V \ldots \prod_{s_i^{C_i}=1}^V \right) 
	\left( p(T_{i(y\underline{s})} = 1 \mid \underline{x_i}, \mathcal{M}, \gamma) \prod_{c}^{C_i} \prod_{a}^{A_i^c} p(r_i^{ca} | s_i^c, \alpha) 
	\right)^{T_{i(y\underline{s})}} \\
\end{align*}

\subsection*{Complete data log-likelihood}

\begin{align*}
\ell(\mathcal{D}, Z | \theta) &=  
	 \sum_{i=1}^N \sum_{y}^V \left( \sum_{s_i^1}^V \ldots \sum_{s_i^{C_i}}^V \right) 
	T_{i(y\underline{s})} \left( 
		\log p(T_{i(y\underline{s})} = 1 \mid \underline{x_i}, \mathcal{M}, \gamma) + \sum_{c}^{C_i} \sum_{a}^{A_i^c} \log p(r_i^{ca} | s_i^c, \alpha) 
	\right) \\
\end{align*}

\subsection*{Expected value of the complete data log-likelihood}


\begin{align*}
\mathbb{E}_Z[\ell(\mathcal{D}, Z | \theta)] &=  
	 \sum_{i=1}^N \sum_{y}^V \left( \sum_{s_i^1}^V \ldots \sum_{s_i^{C_i}}^V \right) 
	\mathbb{E}_Z[T_{i(y\underline{s})}] \left( 
		\log p(T_{i(y\underline{s})} = 1\mid \underline{x_i}, \mathcal{M}, \gamma) + \sum_{c}^{C_i} \sum_{a}^{A_i^c} \log p(r_i^{ca} | s_i^c, \alpha) 
	\right) \\
	 &= \sum_{i=1}^N \sum_{y}^V \left( \sum_{s_i^1}^V \ldots \sum_{s_i^{C_i}}^V \right) 
	\tau_{i(y\underline{s})} \left( 
		\log p(T_{i(y\underline{s})} = 1 \mid \underline{x_i}, \mathcal{M}, \gamma) + \sum_{c}^{C_i} \sum_{a}^{A_i^c} \log p(r_i^{ca} | s_i^c, \alpha) 
	\right) \\
\end{align*}
In the E step, we update the values $\tau$.

\subsection{E step}

Here, we need an expected value for the latent variables conditioned on observed variables and $\theta$. 
Define: 
\begin{align*}
\tau_{i(y\underline{s})} &= p(T_{i(y\underline{s})} = 1 \mid \mathcal{D}, \theta) \\
	&= p(y_i = y, \underline{s_i} = \underline{s} \mid \underline{r_i}, \underline{x_i}, \mathcal{M}, \gamma, \alpha)
\end{align*}

Using Bayes' rule, we have that the update for $\tau_{i(y\underline{s})}$ is:
\begin{align*}
 &= \frac{ p(\underline{r_i} \mid y_i = y, \underline{s_i} = \underline{s}, \underline{x_i}, \mathcal{M}, \gamma, \alpha) p(y_i = y, \underline{s_i} = \underline{s} \mid \underline{x_i}, \mathcal{M}, \gamma, \alpha) } 
 { p(\underline{r_i} \mid \underline{x_i}, \mathcal{M}, \gamma, \alpha) } \\
 &= \frac{ p(y_i = y, \underline{s_i} = \underline{s} \mid \underline{x_i}, \mathcal{M}, \gamma) p(\underline{r_i} \mid \underline{s_i} = \underline{s}, \alpha)  } 
 { p(\underline{r_i} \mid \underline{x_i}, \mathcal{M}, \gamma, \alpha) } \\
 &= \frac{ p(y_i = y \mid \underline{x_i}, \mathcal{M}) p(\underline{s_i} = \underline{s} \mid y_i = y, \gamma) p(\underline{r_i} \mid \underline{s_i} = \underline{s}, \alpha) }  
  { \sum_{y'=1}^V \left(  \sum_{{s'}_i^1}^V \ldots \sum_{{s'}_i^{C_i}}^V \right)  
	p(y_i = y' \mid \underline{x_i}, \mathcal{M}) p(\underline{s_i} = \underline{s'} \mid y_i = y', \gamma) p(\underline{r_i} \mid \underline{s_i} = \underline{s'}, \alpha) }.
\end{align*} 

The numerator is simply the likelihood of a fully observed instance (a tweet and its labels, with the specified setting of latent variables), while the denominator ensures that the distribution sums to 1. 

\subsection{M step for $\gamma$}

Recall that:
\begin{align}
\mathbb{E}_Z[\ell(\mathcal{D}, Z | \theta)] &= 
	 \sum_{i=1}^N \sum_{y}^V \left( \sum_{s_i^1}^V \ldots \sum_{s_i^{C_i}}^V \right) 
	\tau_{i(y\underline{s})} \left( 
		\log p(T_{i(y\underline{s})} \mid \underline{x_i}, \mathcal{M}, \gamma) + \sum_{c}^{C_i} \sum_{a}^{A_i^c} \log p(r_i^{ca} | s_i^c, \alpha) 
	\right) \notag \\
	 &= \sum_{i=1}^N \sum_{y}^V \left( \sum_{s_i^1}^V \ldots \sum_{s_i^{C_i}}^V \right) 
	\tau_{i(y\underline{s})} \left( 
		\log \mathcal{M}_y(\underline{x_i}) + ( \sum_{c}^{C_i} \log {\gamma^c_{ys}} ) + \sum_{c}^{C_i} \sum_{a}^{A_i^c} \log \alpha^a_{sr} 
	\right). \label{expectedComplete}
\end{align}

Recall that $\gamma^c_{ys}$ denotes the matrix entry describing $p(s_i^c = s \mid y_i = y)$. We have a constraint that 
$\sum_{s'}^V \gamma^c_{ys'} = 1$. (That is, within the $c$th matrix of $\gamma$, row $y$ must sum to 1.) Collect terms from $\mathbb{E}_Z[\ell(\mathcal{D}, Z | \theta)]$ that depend on a particular matrix entry $\gamma^c_{ys}$ into one expression $J(\gamma^c_{ys})$, together with the Lagrange multiplier term from the constraint.

\begin{align*}
J(\gamma^c_{ys}) &= 
	\left( 
	\sum_{i=1}^N \sum_{y'}^V ( \sum_{{s'}_i^1}^V \ldots \sum_{{s'}_i^{C_i}}^V ) \sum_{c'}^{C_i}
	\tau_{iy'\underline{s}'} \log {\gamma^{c'}_{y's'}}
	\right)
	- \lambda (\sum_{s'}^V \gamma^c_{ys'} - 1)  \\
	&= 	\left( 
	\sum_{i=1}^N (  \sum_{{s'}_i^1}^V \ldots \sum_{{s'}_i^{C_i}}^V )
		\tau_{i(y\underline{s}')} \log {\gamma^c_{ys'}}
	\right)
	- \lambda (\gamma^c_{ys})  \\
\end{align*}
From the summations over $c'$ and $y'$, only the term with the desired $c$ and $y$ depends on $\gamma^c_{ys}$. In the summations over the values of $\underline{s}$, only the $c$th summation pertains to $\gamma^c_{ys}$ (i.e., $\gamma^c_{ys}$ appears only when ${s'}_i^c = s$). However, the other components of $\underline{s'}$ are latent variables whose probability we need to sum over.

\begin{align*}
	&= 	\left( 
	\sum_{i=1}^N ( \sum_{{s'}_i^1}^V \ldots[\text{except component } c] \ldots \sum_{{s'}_i^{C_i}}^V )
		\tau_{i(y\underline{s}')} \log {\gamma^c_{ys}}
	\right)
	- \lambda (\gamma^c_{ys})  \\
	&= 	 
	\log {\gamma^c_{ys}} \left( \sum_{i=1}^N ( \sum_{{s'}_i^1}^V \ldots[\text{except component } c] \ldots \sum_{{s'}_i^{C_i}}^V )
		\tau_{i(y\underline{s})'} 
	\right)
	- \lambda (\gamma^c_{ys})  \\
	&= 	 
	\log {\gamma^c_{ys}} \left( \text{Weighted number of tweets with $y_i = y$ and $s_i^c = s$} \right)
	- \lambda (\gamma^c_{ys})  \\
\frac{\delta \ell}{\delta \gamma^c_{ys}} = J'(\gamma^c_{ys}) 
	&= \frac{ \text{(Weighted number of tweets with $y_i = y$ and $s_i^c = s$)} } {\gamma^c_{ys}} - \lambda
\end{align*}

Note that ``weighted" always means ``weighted using the current assignment probabilities $\tau_{i(y\underline{s})}$."

Set $J'(\gamma^c_{ys})$ to 0 to get:

\begin{align*}
\gamma^c_{ys} = \frac{ \text{(Weighted number of tweets with $y_i = y$ and $s_i^c = s$)} } { \lambda}.
\end{align*}

Go back to the constraint equation and plug in expression above for each $\gamma$:
\begin{align*}
\sum_{s'}^V \gamma^c_{ys'} = 1 \\
\sum_{s'}^V ( 
	\frac{ \text{(Weighted number of tweets with $y_i = y$ and $s_i^c = s'$)} } { \lambda} ) &= 1 \\
\lambda = (\text{Weighted number of tweets with $y_i = y$ (and any value for $s_i^c$))})
\end{align*}

So,

\begin{align*}
\gamma^c_{ys} = \frac{ \text{(Weighted number of tweets with $y_i = y$ and $s_i^c = s$)} } {(\text{Weighted number of tweets with $y_i = y$ (and any value for $s_i^c$})}.
\end{align*}

\subsection{M step for $\alpha$}

Recall that  $\alpha^a_{sr}$ is the matrix entry describing $p(r_i^{ca} = r \mid s_i^c = s)$---that is, the probability that the $a$th annotator writes $r$ when the tweet (as they saw it in context $c$) had a context-specific label of $s_i^c = s$. Note that each annotation $r_i^{ca}$ takes place in a particular known context $c$, but $\alpha$ does not depend on $c$. To keep track---while we move terms around---of the context associated with each annotation, we re-expand $\log p(r_i^{ca} | s_i^c, \alpha)$ to $\alpha^a_{sr} \delta(s_i^c = s)$.  

Referring back to Eq. \eqref{expectedComplete}, collect terms that depend on $\alpha^a_{sr}$. Also add the normalization constraint that $\sum_{r'}^V \alpha^a_{sr'} = 1$.

\begin{align*}
\mathbb{E}_T[\ell(\mathcal{D}, Z | \theta)]
	 &= \sum_{i=1}^N \sum_{y}^V \left( \sum_{s_i^1}^V \ldots \sum_{s_i^{C_i}}^V \right) 
	\tau_{iy\underline{s}} \left( 
		\log \mathcal{M}_y(\underline{x_i}) + ( \sum_{c}^{C_i} \log {\gamma^c_{ys}} ) + \sum_{c}^{C_i} \sum_{a}^{A_i^c} \delta(s_i^c = s) \log \alpha^a_{sr} 
	\right) \\
J(\alpha^a_{sr}) &= \left( \sum_{i=1}^N \sum_{y}^V ( \sum_{{s'}_i^1}^V \ldots \sum_{{s'}_i^{C_i}}^V )
	   \sum_{c}^{C_i} \sum_{a'}^{A_i^c} \tau_{i(y\underline{s})} \delta(s_i^c = s) \log \alpha^{a'}_{sr} \right) - \lambda (\sum_{r'}^V \alpha^a_{sr'} - 1) \\
	 &= \left( \sum_{i=1}^N \sum_{y}^V \sum_{c}^{C_i} ( \sum_{{s'}_i^1}^V \ldots \sum_{{s'}_i^{C_i}}^V ) 
	 \tau_{i(y\underline{s})} \delta(s_i^c = s) \log \alpha^a_{sr} \right) - \lambda (\alpha^a_{sr}) \\
	 &= \left( \sum_{i=1}^N \sum_{y}^V \sum_{c}^{C_i} ( \sum_{{s'}_i^1}^V \ldots[\text{except component } c] \ldots \sum_{{s'}_i^{C_i}}^V ) 
	  \tau_{i(y\underline{s})}  \log \alpha^a_{sr} \right) - \lambda (\alpha^a_{sr}) \\
	 &= \log \alpha^a_{sr}  \left(  \sum_{i=1}^N \sum_{y}^V \sum_{c}^{C_i} (\sum_{{s'}_i^1}^V \ldots[\text{except component } c] \ldots \sum_{{s'}_i^{C_i}}^V  )
	 \tau_{i(y\underline{s})} \right) - \lambda (\alpha^a_{sr}) \\
	 &= \log \alpha^a_{sr}   \text{(Weighted number of annotations with value $r$ by annotator $a$ having $s_i^c = s$)} - \lambda (\alpha^a_{sr}) \\
\end{align*}
\begin{align*}	 
\frac{\delta \ell}{\delta \alpha^a_{sr}} = J'(\alpha^a_{sr}) &= 
	 \frac{ \text{(Weighted number of annotations with value $r$ by annotator $a$ having $s_i^c = s$)} } { \alpha^a_{sr} } - \lambda
\end{align*}

Notice that for $\gamma$, we were counting tweets in a particular context (and looking at their configuration of latent variables). For $\alpha$ here, the sum is over tweets + contexts; we are counting all annotations made by a particular annotator (and looking at the $s_i$ for the context in which the annotation took place).

Set $J'(\alpha^a_{sr})$ to 0 to get:
\begin{align*}
\alpha^a_{sr} = \frac{(\text{Weighted number of annotations with value $r$ by annotator $a$ in which $s_i^c = s$})}{\lambda}.
\end{align*}

The constraint equation works just like it did for $\gamma$:

\begin{align*}
\sum_{r'}^V \alpha^a_{sr'} = 1 \\
\frac{1}{\lambda} \sum_{r'}^V (\text{Weighted number of annotations with value $r'$ by annotator $a$ in which $s_i^c = s$}) &= 1 \\
\lambda = ( \text{Weighted number of annotations by annotator $a$ in which $s_i^c = s$} ).
\end{align*}

Finally,

\begin{align*}
\alpha^a_{sr} = \frac{(\text{Weighted number of annotations with value $r$ by annotator $a$ in which $s_i^c = s$})} {\text{(Weighted number of annotations by annotator $a$ in which $s_i^c = s$)}}.
\end{align*}

\subsection{Computing labels to use for classifiers} \label{sec:class}

Using Raykar et al.'s suggestion, we decide to test a variety of classifiers. In order to do so, we must recover $\mathbb{E}_Z[y_{i} = k]$, the expected likelihood of $y_i$ taking on the particular value $k$. Starting from $\tau_{i(k\underline{s})}$, which is computed during the E step and  is defined as $p(y_i=k,\underline{s_i} = \underline{s} | \mathcal{D}, \theta)$, we marginalize out $\underline{s_i}$:

\begin{align*}
\mathbb{E}_Z[y_{i} = k] = p(y_i=k \mid \mathcal{D}, \theta) =  \sum_{s_i^1}^V \ldots \sum_{s_i^{C_i}}^V \tau_{i(k\underline{s})}
\end{align*}

We can then use these probability values to train any multi-class classifier we wish by performing sampling based on the obtained weights.  During model testing and evaluation, we observed that the number of samples per item did not significantly impact model performance. Therefore, for all results presented in the paper, we simply fixed the number of samples per item to 10.

\section{EM Algorithm Debugging}\label{sec:deb}

In order to ensure the algorithm, as coded, correctly learns parameters, we take two steps.  First, we ensure that the log-likelihood of the model decreases on every iteration.  Second, we developed simulations to ensure that we can recover known parameters for simulated data.  Simulations suggested that the model was easily able to uncover known parameters for $\gamma$ across a variety of tested values and conditions similar to those that generated our data (i.e. with the same numbers of tweets, context conditions and annotators).  However, we observe that the model does struggle to recover some parameterizations of $\alpha$; we expect the cause of this is a combination of the randomness induced by the data generating process and the sheer number of $\alpha$ parameters in the model ($9|A|$).  Future work might consider how to limit the number of parameters in $\alpha$ by, e.g., assuming annotators are a mixture over a smaller number of prototypical annotation styles.

%
%

\section{Hyperparameter Optimization}\label{sec:hyp}

For all hyperparameter tuning, we use a rough grid search approach, testing various settings on performance on the development set  (focusing on both Log-Loss and Average F1). As our intention was to focus on the impact of different labeling schemes, our goal in hyperparameter tuning was simply to find reasonable and, more importantly, consistent models that we could use to address the impact of the labeling structure (and our model ablations). As noted in the paper, future work will focus on improving our model and 

For hyperparameter tuning of the baseline models, we tune parameters for the maximum depth of the tree and the number of estimators. For our model, our model ablations and each of the baselines, a maximum depth of 30 and 3000 estimators were used, as results stabilized around these numbers.  

Development and validation differed in that the development data consisted of only registered Democrats and Republicans, while the   validation data also generalized to non-labeled Democrats and Republicans. Because of this, we allowed the baseline models to ``cheat'' by setting class weights for the Random Forest to the ratio of true label counts in the validation data versus the training data. Note we did \emph{not} do this for ConStance or its ablations, setting the prior on $y$ via tuning on the test set. 

To tune ConStance and the tested ablations, we considered varying the initializations of $\alpha$ and $\gamma$, providing Dirichlet priors on $\alpha$ and $\gamma$ and by varying a prior on $y$. In the end, initialization of $\alpha$ and $\gamma$ made little difference, we elected not to provide the model with any priors on $\alpha$ or $\gamma$. However, we did set the prior on $y$ to $[.495,.01,.495]$ for the ``Trump'', ``Neutral'' and ``Clinton'' labels, respectively.  Like the class weights for the baseline Random Forest models, this prior urged the model away from selecting the ``Neutral'' option, which was far less prevalent in the test and validation data than it was in the annotations. This, of course, is because annotating with full context allowed for a significantly more discriminative take on the support of Twitter users as compared to the context seen by the AMT workers.

For the ablations, models performed better with a different prior on $y$ ($[.45,.1,.45]$), we therefore use this for validation.

\end{document}